\documentclass[aps,prl,superscriptaddress,showpacs,floatfix,amsmath,amssymb,nofootinbib,preprintnumbers,
twocolumn]{revtex4-1}

\usepackage{graphicx}

\usepackage{epstopdf}

\usepackage{color}

\newcommand{\be}{\begin{equation}}
\newcommand{\ee}{\end{equation}}
\newcommand{\ba}{\begin{eqnarray}}
\newcommand{\ea}{\end{eqnarray}}

\newcommand{\beq}{\begin{equation}}
\newcommand{\eeq}{\end{equation}}
\newcommand{\beqa}{\begin{eqnarray}}
\newcommand{\eeqa}{\end{eqnarray}}

\newcommand{\bea}{\begin{eqnarray}}
\newcommand{\eea}{\end{eqnarray}}

\begin{document}
\title{Isolated critical point from Lovelock gravity}

\author{Brian P. Dolan}
\email{B.P.Dolan@hw.ac.uk}
\affiliation{Department of Mathematics, Heriot-Watt University,
Colin Maclaurin Building, Riccarton, Edinburgh, EH14 4AS, U.K.}
\affiliation{Maxwell Institute for Mathematical Sciences, Edinburgh, U.K.}
\author{Anna Kostouki}
\email{akostouki@perimeterinstitute.ca}
\affiliation{Perimeter Institute, 31 Caroline St. N., Waterloo,
Ontario, N2L 2Y5, Canada}
\author{David Kubiz\v n\'ak}
\email{dkubiznak@perimeterinstitute.ca}
\affiliation{Perimeter Institute, 31 Caroline St. N., Waterloo,
Ontario, N2L 2Y5, Canada}
\affiliation{Department of Physics and Astronomy, University of Waterloo,
Waterloo, Ontario, Canada, N2L 3G1}
\author{Robert B. Mann}
\email{rbmann@uwaterloo.ca}
\affiliation{Department of Physics and Astronomy, University of Waterloo,
Waterloo, Ontario, Canada, N2L 3G1}

\date{July 17, 2014}  

\begin{abstract}
For any $K(=2k+1)$th-order Lovelock gravity with fine-tuned Lovelock couplings, we demonstrate the existence of a special isolated critical point characterized by non-standard critical exponents in the phase diagram of hyperbolic vacuum black holes.  In the Gibbs free energy this corresponds to a place wherefrom two swallowtails emerge, giving rise to two first-order phase transitions between small and large black holes. We believe that this is a first example of a critical point with non-standard critical exponents obtained in a geometric theory of gravity. 
\end{abstract}

\pacs{04.50.Gh, 04.70.-s, 05.70.Ce}

\maketitle

 
The thermodynamic behaviour of asymptotically anti de Sitter (AdS) black holes has been a subject of interest ever since the discovery of thermal radiation/large AdS black hole phase transitions \cite{HawkingPage:1983}.  Not only is it straightforward to define thermodynamic equilibrium, but 
these spacetimes also  admit a gauge duality description via a dual thermal field theory, providing important insight into the underlying structure of quantum gravity.

A new development in this area occurred with the proposal that the cosmological constant be interpreted as a thermodynamic variable \cite{CreightonMann:1995}  that plays the role of pressure
\cite{CaldarelliEtal:2000, KastorEtal:2009, Dolan:2010}.  This in turn implies that the
 mass of the black hole should be regarded as the {\it enthalpy} of spacetime: 
namely the sum of  both its internal energy  and the energy  required to ``make room for it" by displacing its (vacuum energy) environment.  This perspective has led to a number of novel insights and new phenomena in black hole thermodynamics, including the  realization that the phase-transition between large AdS black holes and radiation \cite{HawkingPage:1983} can be understood as a ``liquid/solid'' phase transition \cite{Kubiznak:2014zwa}, the discovery that charged black holes behave as Van der Waals fluids \cite{ChamblinEtal:1999a, Cvetic:1999ne, KubiznakMann:2012}, 
and the discoveries of  {\em reentrant phase transitions} \cite{Altamirano:2013ane}, in which 
there are phase transitions from large  black holes to small ones and then back to large again
as the temperature monotonically increases, and of 
 {\em triple points} 
 \cite{Altamirano:2013uqa} for  Kerr-AdS black holes, where  a coalescence of
small, medium, and large sized black holes merge into a single kind at a particular critical value of the pressure and temperature, analogous to the triple point of water.  

Understanding the nature of critical points for these phenomena provides additional insight into black hole
thermodynamics  and may reveal key insights into  quantum gravity. So far all black holes studied have been shown to have the standard set of critical exponents expected from
mean field theory: $\alpha=0$, $\beta=\frac{1}{2}$, $\gamma=1$ and $\delta=3$.  
However, it is well known that many statistical systems demonstrate critical exponents different from these, the Ising model in less than 4 dimensions being the best known example.
Here we report on the first known case of black holes that do not have mean-field critical exponents.  We find that they occur in all  Lovelock gravity theories of odd order $K$.

Non-mean-field exponents are normally interpreted as an indication that long distance interactions of some kind are important in the microscopic physics; however scaling arguments usually imply that the exponents should be mean-field in dimensions greater than 4.  Here we find non mean-field exponents in any dimension greater than or equal to 5.  Furthermore, these are determined not by summing over states in a microscopic description of any specific dimension, but rather from a single classical solution of a gravitational theory.

Lovelock gravity \cite{Lovelock:1971yv} is a geometric higher curvature theory of gravity that can be considered as a natural generalization of Einstein's theory to higher dimensions---it is the unique higher-derivative theory that gives rise to second-order field equations for all metric components.  Its Lagrangian is  \cite{Lovelock:1971yv}
 \begin{equation}
\mathcal{L}=\frac{1}{16\pi G_N}\sum_{k=0}^{K}\hat{\alpha}_{\left(k\right)}\mathcal{L}^{\left(k\right)}\,, 
\label{eq:Lagrangian}
\end{equation}
in  $d$ spacetime dimensions, 
 where $K=\left[\frac{d-1}{2}\right]$, the  $\hat{\alpha}_{\left(k\right)}$ are the  Lovelock coupling constants, and $\mathcal{L}^{\text{\ensuremath{\left(k\right)}}}$ are the $2k$-dimensional Euler densities, given by 
$\mathcal{L}^{\left(k\right)}=\frac{1}{2^{k}}\,\delta_{c_{1}d_{1}\ldots c_{k}d_{k}}^{a_{1}b_{1}\ldots a_{k}b_{k}}R_{a_{1}b_{1}}^{\quad c_{1}d_{1}}\ldots R_{a_{k}b_{k}}^{\quad c_{k}d_{k}}\,,
$
with the  `generalized Kronecker delta function' $\delta_{c_{1}d_{1}\ldots c_{k}d_{k}}^{a_{1}b_{1}\ldots a_{k}b_{k}}$  totally antisymmetric in both sets of indices, and $R_{a_{k}b_{k}}^{\quad c_{k}d_{k}}$  the Riemann tensor. 
In what follows we always take all the Lovelock couplings to be positive and 
identify the (negative) cosmological constant $\Lambda=-\hat{\alpha}_{0}/2$ with  thermodynamic pressure  \cite{CaldarelliEtal:2000, KastorEtal:2009, Dolan:2010}
\begin{equation}\label{P}
P=-\frac{\Lambda}{8\pi G_N} = \frac{\hat{\alpha}_{0}}{16\pi G_N}\,,
\end{equation}
allowing it to vary in the first law of black hole thermodynamics. We associate the conjugate quantity to $P$ as the thermodynamic volume $V$. With this identification, the mass $M$ of the black hole is interpreted as   enthalpy rather than internal energy \cite{KastorEtal:2009}.

In what follows we concentrate on static vacuum spherically symmetric AdS Lovelock black holes with hyperbolic horizon geometry, employing the ansatz  
\be\label{solution}
ds^{2} = -f\left(r\right)dt^{2}+f\left(r\right)^{-1}dr^{2}+r^{2}d\Omega_{d-2}^{2}\,,
\ee
where $d\Omega_{d-2}^{2}$ denotes the line element of a $\left( d-2 \right)$-dimensional  space of constant curvature $\kappa(d-2)(d-3)$, with  $\kappa=+1,0,-1$ for spherical, flat, and hyperbolic  
geometries respectively of finite   volume   $\Sigma_{d-2}$, the latter two cases being compact via identification
\cite{Aminneborg:1996iz,Smith:1997wx,Mann:1997iz}.

 The Lovelock equations from \eqref{eq:Lagrangian} reduce after integration to the following polynomial equation for $f$ 
\cite{wheeler1986symmetric2, Cai:2003kt}:
\begin{equation} \label{eq:poly}
{\cal P}\left(f\right)=\sum_{k=0}^{K}\alpha_{k} \left(\frac{\kappa-f}{r^2}\right)^{k}=\frac{16\pi G_NM}{(d-2)\Sigma_{d-2}r^{d-1}}\equiv m(r)\,,
\end{equation}
where $M$ stands for the ADM mass of the black hole, and
\ba
\alpha_{0}&=&\frac{\hat{\alpha}_{(0)}}{\left(d-1\right)\left(d-2\right)}=\frac{16\pi G_N P}{\left(d-1\right)\left(d-2\right)}\,,
\quad{\alpha}_{1}={\hat \alpha}_{(1)}\,,\nonumber\\
\alpha_{k}&=&\hat \alpha_{(k)}\prod_{n=3}^{2k}\left(d-n\right)  {\quad\mbox{for}\quad  k\geq2}
\ea
are the rescaled Lovelock couplings.

We consider a very special case of Lovelock gravity  such that ($\alpha\equiv \alpha_K)$
\be\label{luvpoly}
{\cal P}\left(f\right)=\alpha\left(\frac{\kappa-f}{r^2}+A\right)^K - \alpha A^K 
+\alpha_0\,.
\ee 
which implies\footnote{Note there is a certain similarity with a class of Chern--Simons theories \cite{Zanelli:2005sa}, for which the 
Lovelock couplings obey (with $d$ odd)
\be
\alpha_p =\frac{\ell^{2p-2n+1}}{2n-2p-1}\left( {n-1\atop p}  \right)  \qquad p=1,2,\ldots, n-1 = \frac{d-1}{2}\,,
\ee
with $\ell$ being the AdS radius.}
\be
 \alpha_k =  \alpha A^{K-k}\left( {K\atop k}\right)\,, \qquad 2\leq k<K\,,
\ee  
with  $\alpha_0$ arbitrary and $\alpha_1=1$. The requirement  (\ref{luvpoly}) also implies $A=(K\alpha)^\frac{-1}{K-1}$ and yields
\be
f=\kappa+r^2A\bigg[1-\Bigl(\frac{m(r)-\alpha_0}{\alpha A^K}+1\Bigr)^{1/K}\bigg]\,.
\ee 
The black hole mass $M$, the temperature $T$, the entropy $S$, the thermodynamic volume $V$, and the potentials $\Psi$ conjugate to $\alpha$ read
($B\equiv \frac{\kappa}{r_+^2}+A$)
\bea
M&=&\frac{(d-2)\Sigma_{d-2}r_+^{d-1}}{16\pi G_N}\bigl(\alpha B^K-\alpha A^K+\alpha_0\bigr)\,,\\
T &=&  \frac{\vert f^\prime(r_+)\vert}{4\pi} =\frac{-\kappa}{2\pi r_+}+\frac{(d-1)r_+}{4\pi K B^{K-1}}\Bigl(B^K-A^K+\frac{\alpha_0}{\alpha}\Bigr)\,,\quad  \label{T}\nonumber\\
S&=&\frac{\Sigma_{d-2}\left(d-2\right)\alpha}{4G_N}\sum_{k=1}^{K}\left( {K\atop k}\right)\frac{k\kappa^{k-1}A^{K-k}r_+^{d-2k}}{d-2k}\,, \nonumber\\
V&=&\frac{\Sigma_{d-2}r_+^{d-1}}{d-1}\,, \quad \Psi=\sum_{k=2}^K \frac{k-1}{K-1}\psi^{(k)}A^{K-k}\left( {K\atop k}  \right)\,,\label{V}
\eea
where
\be 
\psi^{(k)}=\frac{\Sigma_{d-2}(d-2)}{16\pi G_N}\kappa^{k-1}{r^{d-2k}_+}\left[\frac{\kappa}{r}-\frac{4\pi kT}{d-2k}\right]\,, \quad k\geq 2\,.
\ee
All these quantities satisfy the first law of black hole thermodynamics and the corresponding Smarr--Gibbs--Duhem relation
\ba
\delta M&=&T\delta S+V\delta P+\Psi\delta \alpha\,,\label{first}\nonumber\\
\left(d-3\right)M&=&\left(d-2\right)TS-2 VP+2\left(K-1\right)\Psi\alpha\,.\qquad\label{Smarr}
\ea

Re-arranging   \eqref{T}, we have the following equation of state:
\be
P=\frac{(d\!-\!1)(d\!-\!2)\alpha}{16\pi G_N}\bigg[B^{K\!-\!1}\Bigl(\frac{2K(2\pi r_+T\!+\!\kappa)}{(d-1)r_+^2}
-B\Bigr)+A^K\bigg]\,,
\ee 
where $r_+=r_+(V)$ through the relation \eqref{V}. Alternatively we can define a `specific volume', $v=\frac{V}{N} \frac{4(d-1)}{d-2}$, where $N$ measures the number of degrees of freedom associated with the black hole horizon.  In Einstein gravity $N=A/L_{\mbox{\tiny Planck}}^2$, and the dimension-dependent factor $\frac{4(d-1)}{d-2}$ is chosen such that the equation of state takes the ideal gas law, $P=\frac{T}{v}+\dots$, to leading order.
For simplicity, we shall use $r_+=r_+(V)$ and in what follows   concentrate on the $\kappa=-1$ case.

\begin{figure*}
\centering
\begin{tabular}{cc}
\rotatebox{-90}{
\includegraphics[width=0.34\textwidth,height=0.28\textheight]{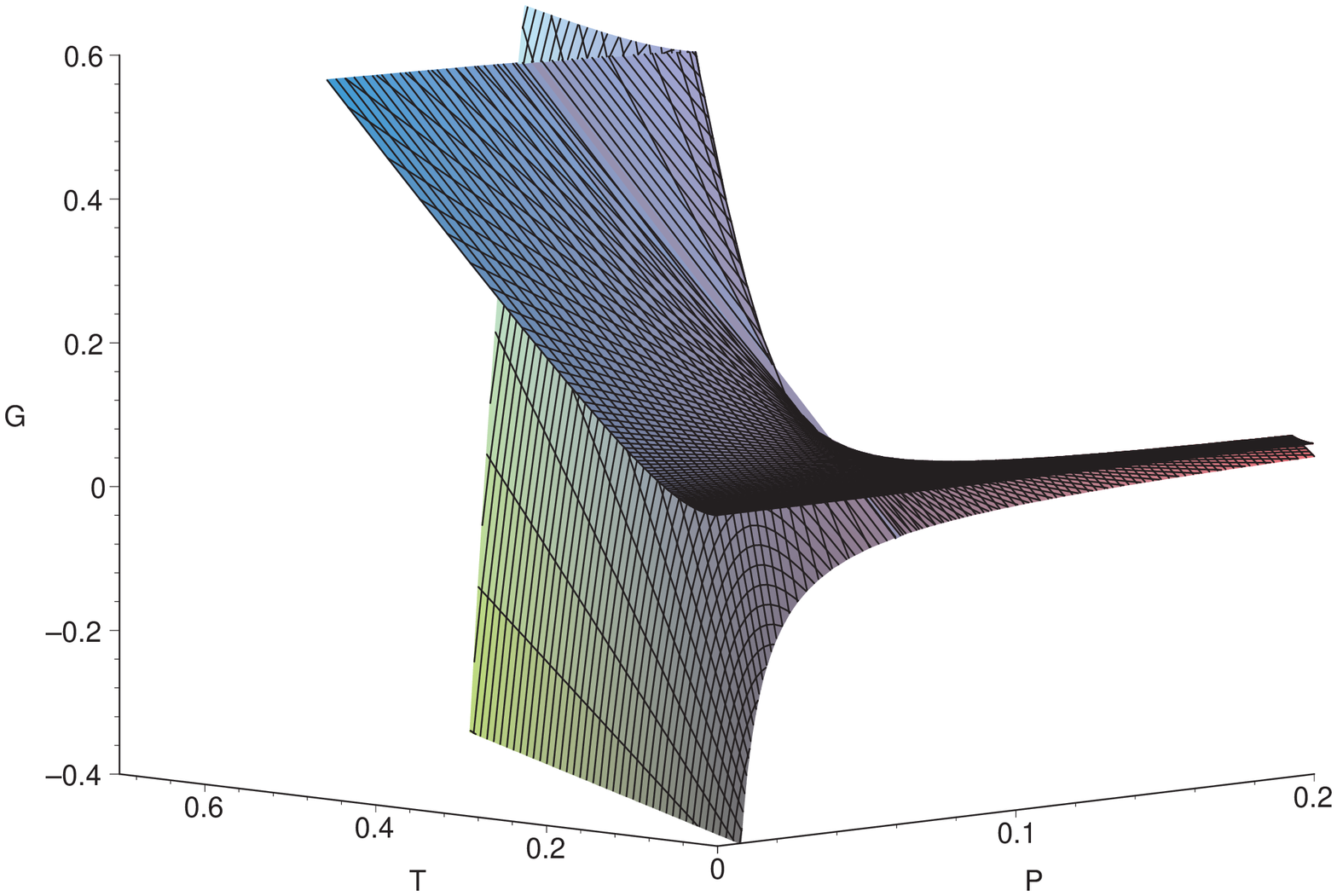}} &
\rotatebox{-90}{
\includegraphics[width=0.34\textwidth,height=0.28\textheight]{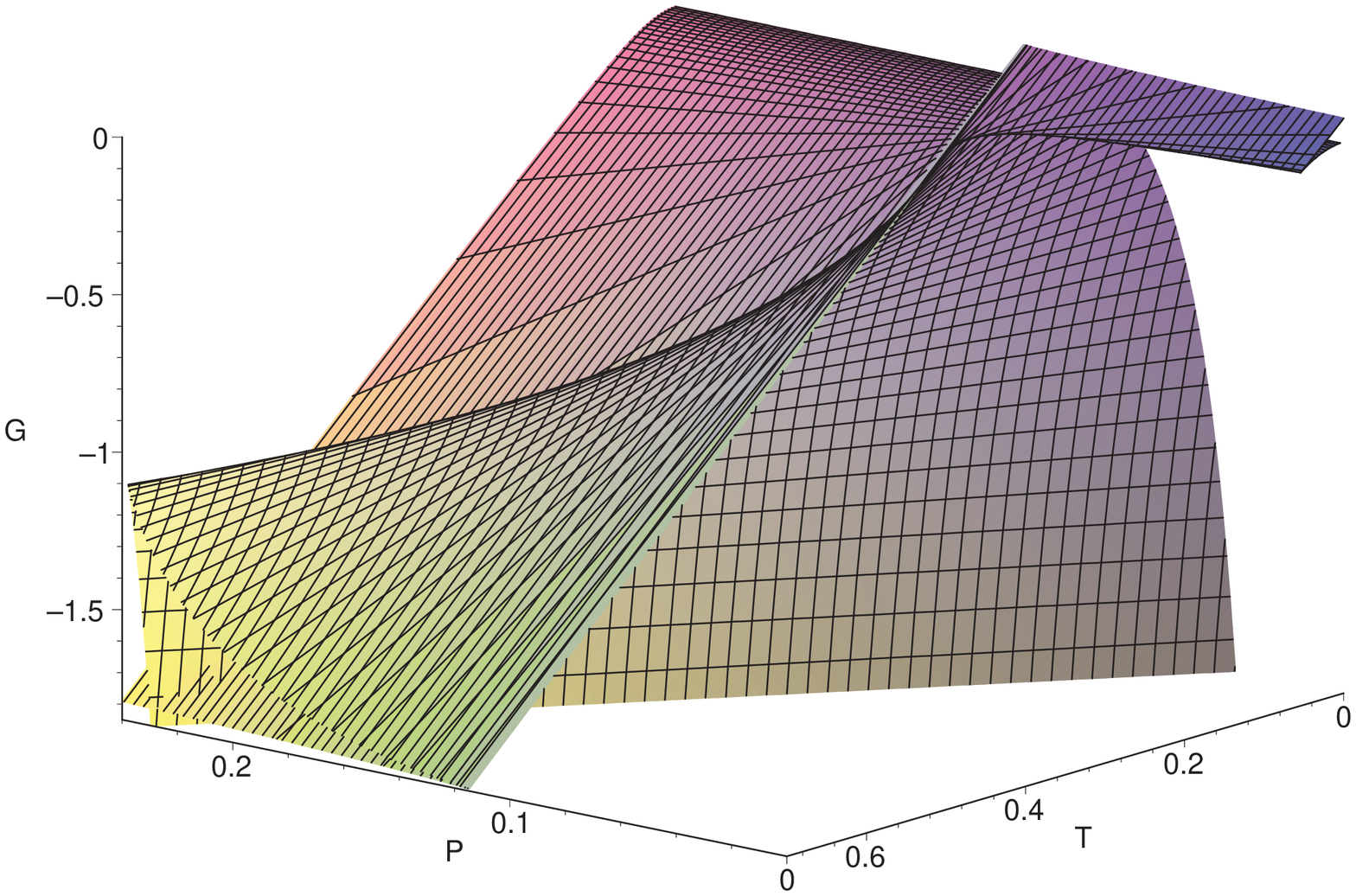}}\\
\end{tabular}
\caption{{\bf Gibbs free energy.}
{\em Left:} The Gibbs free energy is displayed for $K=2$ and $d=5$--- there is no criticality associated with the special point \eqref{crit}. 
For all even-oder Lovelock black holes in an arbitrary dimension the  behaviour is qualitatively similar. {\em Right: }The Gibbs 
free energy is displayed for $K=3$ and $d=7$ dimensions. We now observe two swallow tails and the associated first-order phase transitions, both emanating from the same (isolated) critical point. Qualitatively same situation occurs for all odd-order Lovelock black holes in higher dimensions. 
In both plots we have set $\Sigma_d=1\,,\alpha=1\,,G_N=1$.   
}  
\label{fig}
\end{figure*}

To analyze possible critical points of this equation we compute the derivatives of $P$ with respect to $r_+$. We find that $\frac{\partial^k P}{\partial r_+^k}=0$ for all $k=1,\dots, K-2$ provided that $r_+=r_c=1/\sqrt{A}$. We can also arrange that $\frac{\partial^{K-1} P}{\partial r_+^{K-1}}=0$ provided we fix $T=T_c=(2\pi r_c)^{-1}$. Finally, we find that $\frac{\partial^{K} P}{\partial r_+^{K}}$ is always negative. This means that we have a special point, given by 
\be\label{crit}
r_c=\frac{1}{\sqrt{A}}\,,\quad T_c=\frac{1}{2\pi r_c}\,,\quad P_c=\frac{(d-1)(d-2)\alpha}{16\pi G_N} A^K\,.
\ee
When $K$ is even this point corresponds to a maximum of $P$ and there is no associated criticality (a fact further confirmed by examining  the Gibbs free energy, shown in Fig.~\ref{fig}). However, when $K$ is odd, the special point is a point of inflection of a   strictly decreasing function, and so describes an isolated critical point where the two first-order phase transitions merge (or in terms of the Gibbs free energy, the two swallowtails merge).   Note that at this point the mass of the black hole vanishes, $M=0$.

Introducing the new variables
\be
\omega=\frac{r_+}{r_c}-1\,,\quad \tau=\frac{T}{T_c}-1\,,
\ee
we find the following expansion near the critical point:
\be
\frac{P}{P_c}=1+K\frac{2^K}{d-1}\omega^{K-1}\tau 
+\frac{(K-d+1)2^K}{d-1}\omega^K+\dots
\ee
From this expansion it follows that we have the following critical exponents:
\be
\tilde \beta=1\,,\quad \tilde \gamma=K-1\,,\quad \tilde \delta=K\,.
\ee
We also find that identically $C_V=0$, implying that 
\be
\tilde \alpha=0\,.
\ee

Although non-standard, these critical exponents satisfy both the Widom relation
\be
\tilde \gamma=\tilde \beta(\tilde \delta-1)\,,
\ee
and the Rushbrooke inequality 
\be
\tilde \alpha+2\tilde \beta+\tilde \gamma\geq 2\,. 
\ee
We obtain a strict inequality because $C_P \sim |\tau|^{K-1}$
vanishes at the critical point. 

The thermodynamic potential to consider in the canonical ensemble is the Gibbs free energy, given by \cite{Kastor:2010gq} (see also  \cite{Kofinas:2007ns} for the Euclidean action calculation) 
\be\label{G}
G=M-TS=G(P,T, \alpha)\,.
\ee
The thermodynamic state corresponds to the global minimum of this quantity for   fixed parameters $P, T$ and $\alpha$.
The behavior of $G$ depends crucially on the order of Lovelock gravity. Namely, for $K$ odd we observe two swallow tails starting from the same critical point, whereas in even-order Lovelock,  there is no criticality associated with this point, as shown in Fig.~\ref{fig}.
 
We find that the branches of black holes that globally minimize the Gibbs free energy (and possess non-negative temperature) have always non-negative $C_P$ and hence are locally thermodynamically stable. At the critical point we have $C_P=0$, as we must to violate the Rushbrooke equality as discussed above.

\begin{figure}
\begin{center}
\rotatebox{-90}{
\includegraphics[width=0.34\textwidth,height=0.31\textheight]{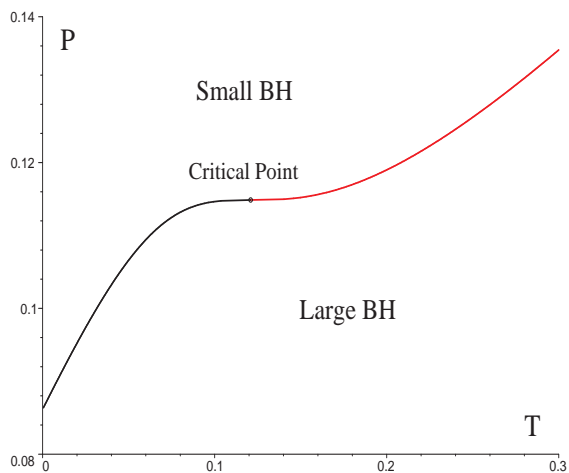}
}
\caption{{\bf Phase diagram.}
The $P-T$ phase displays two phases of black holes: large and small, separated by two first order-phase transitions (with coexistence lines denoted by black and red curves) that both emerge from a single isolated critical point where the phase transition becomes continuous. 
We have displayed $K=3, d=7$ case, higher $d$ and higher (odd) $K$ behave qualitatively similar.
}  
\label{Fig2}
\end{center}
\end{figure}

The $P-T$ phase diagram is displayed in Fig.~\ref{Fig2}. We observe two coexistence lines of the first-order small/large black hole phase transition that both terminate at the isolated critical point \eqref{crit}. At this point the phase transition becomes continuous 
and occurs for massless, $M=0$, black holes.  


To summarize, we have found the first example of an isolated critical point in a geometric theory of gravity,
generalizing results recently obtained in a more specific context  \cite{Mo:2014qsa, Xu:2014tja, Frassino:2014pha}.    The odd-order
Lovelock theories (in any dimension in which they exist) all have massless topological black holes \cite{Smith:1997wx,Mann:1997iz}  as solutions, and so will exhibit this phenomenon for an appropriate choice of coupling constants.  It is straightforward to check that the Ehrenfest equations are 
trivially satisfied at this isolated critical point (both sides identically vanish).  We have also computed Prigogine--Defay ratio \cite{Prigo:book} to be $\Pi=1/K$ for this class of black holes, indicating that the phase transition has more than one order parameter. Our results suggest that  the microscopic degrees of freedom for this class of black holes have 
strong correlations, e.g., \cite{gundermann2011predicting}, whose origin remains to be understood.

\vskip -1.0in
\section*{Acknowledgments} 
This research was supported in part by Perimeter Institute for Theoretical Physics and by the Natural Sciences and Engineering Research Council of Canada. Research at Perimeter Institute is supported by the Government of Canada through Industry Canada and by the Province of Ontario through the Ministry of Research and Innovation.
\vskip -0.25in


\providecommand{\href}[2]{#2}\begingroup\raggedright\endgroup

\end{document}